\documentclass[12pt]{article}
 \usepackage{epsfig}
 \def\be{\begin{equation}}
 \def\ee{\end{equation}}
 \def\bea{\begin{eqnarray}}
 \def\eea{\end{eqnarray}}
 \usepackage{graphicx}

 \catcode`\@=11
 \def\lsim{\mathrel{\mathpalette\@versim<}}
 \def\gsim{\mathrel{\mathpalette\@versim>}}
 \def\@versim#1#2{\vcenter{\offinterlineskip
 \ialign{$\m@th#1\hfil##\hfil$\crcr#2\crcr\sim\crcr } }}
 \catcode`\@=12

 \catcode`@=12
\begin{document}
 \thispagestyle{empty}
 \begin{flushright}
 UCRHEP-T609\\
 Feb 2021\
 \end{flushright}
 \vspace{0.6in}
 \begin{center}
 {\LARGE \bf Dirac Neutrino Mass Matrix and\\ 
 its Link to Freeze-in Dark Matter\\}
 \vspace{1.2in}
 {\bf Ernest Ma\\}
 \vspace{0.2in}
{\sl Physics and Astronomy Department,\\ 
University of California, Riverside, California 92521, USA\\}
\end{center}
 \vspace{1.2in}

\begin{abstract}\
Using a mechanism which allows naturally small Dirac neutrino masses and its 
linkage to a dark gauge $U(1)_D$ symmetry, a realistic Dirac neutrino mass 
matrix is derived from $S_3$.  The dark sector naturally contains a 
fermion singlet having a small seesaw mass.  It is thus a good candidate for 
freeze-in dark matter from the decay of the $U(1)_D$ Higgs boson.
\end{abstract}

 \newpage
 \baselineskip 24pt
\noindent \underline{\it Introduction}~:~
It has been shown recently~\cite{m21} that naturally small Dirac neutrino 
masses may be linked to a dark $U(1)_D$ gauge symmetry.  One specific model 
is studied here with the inclusion of an $S_3$ family symmetry, so that 
a realistic Dirac neutrino mass matrix is obtained. 
The dark sector consists of four singlet Majorana fermions.  Its structure 
allows one to be the lightest from a seesaw mechanism akin to that used 
in canonical Majorana neutrino mass.  It is thus very suitable as freeze-in 
dark matter which owes its relic abundance from the decay of the $U(1)_D$ 
Higgs boson.

The simple mechanism in question was first pointed out in 2001~\cite{m01}.  
Consider two Higgs dooublets $\Phi = (\phi^+,\phi^0)$ and 
$\eta = (\eta^+,\eta^0)$, where $\eta$ is distinguished from the 
standard-model (SM) $\Phi$ by a symmetry to be decided.  Whereas $\Phi$ 
has the usual $\mu^2 < 0$, the corresponding $m^2$ for $\eta$ is positive 
and large.  The aforesaid symmetry is assumed to be broken by the soft term 
${\mu'}^2 \Phi^\dagger \eta + H.c.$  The spontaneous breaking of the  
$SU(2)_L \times U(1)_Y$ gauge symmetry of the SM then results in the usual 
vacuum expectation $\langle \phi^0 \rangle = v$, but 
$\langle \eta^0 \rangle = v'$ is now given by $-{\mu'}^2 v/m^2$, which is 
suppressed by the small ${\mu'}^2$ and large $m^2$.

For neutrino mass, if $\nu_R$ is chosen to transform in the same way as 
$\eta$, but not the other SM particles, then it pairs up with $\nu_L$ to 
form a Dirac fermion with mass proportional to the small $v'$.  If the 
symmetry chosen also forbids $\nu_R$ to have a Majorana mass, then the 
neutrino is a Dirac fermion with a naturally small mass.  This idea of 
achieving a small $v'/v$ ratio is akin to that of the so-called Type II  
seesaw, as classified in Ref.~\cite{m98} and explained in Ref.~\cite{ms98}.  
It is also easily generalized~\cite{glr09} and  applicable to light quarks 
and charged leptons~\cite{m16}.

In this paper, following Ref.~\cite{m21} which incorporates an anomaly-free 
$U(1)_D$ gauge symmetry to distinguish $\nu_R$ from the other SM particles, 
a specific model of two massive Dirac neutrinos is proposed.  With the 
implementation of an $S_3$ discrete family symmetry, a realistic Dirac 
neutrino mass matrix is obtained.  The natural occurrence of light 
freeze-in dark matter is also discussed.

\noindent \underline{\it Outline of Model}~:~
The particle content of the proposed model is listed in Table 1.
\begin{table}[tbh]
\centering
\begin{tabular}{|c|c|c|c|c|}
\hline
fermion/scalar & $SU(2)_L$ & $U(1)_Y$ & $U(1)_D$ & $S_3$ \\
\hline
$(\nu,e),[(\nu_\mu,\mu),(\nu_\tau,\tau)]$ & 2 & $-1/2$ & 0 & 
$1',2$ \\ 
$e^c,[\tau^c,\mu^c]$ & 1 & 1 & 0 & $1',2$ \\
\hline
$\nu^c_{2,3}$ & 1 & 0 & $-4$ & $1,1'$ \\ 
$\psi_{1,2,3}$ & 1 & 0 & 1 & $1,1,1$ \\
$\zeta$ & 1 & 0 & 5 & $1$ \\ 
\hline
\hline
$\Phi_{1,2}=(\phi_{1,2}^+,\phi_{1,2}^0)$ & 2 & 1/2 & 0 & $1,1'$ \\ 
\hline
$\eta_{1,2,3,4}=(\eta_{1,2,3,4}^+,\eta_{1,2,3,4}^0)$ & 2 & 1/2 & $4$ & 
$1,1',2$ \\ 
$\chi_{1}$ & 1 & 0 & 2 & $1$ \\ 
$\chi_2$ & 1 & 0 & 4 & 1 \\
$\chi_3$ & 1 & 0 & 6 & 1 \\
\hline
\end{tabular}
\caption{Fermion and scalar content of Dirac neutrino model with dark 
$U(1)_D$ and $S_3$ symmetries.}
\end{table}

The $U(1)_D$ gauge symmetry is anomaly-free becaue
\begin{equation}
1+1+1-4-4+5=0, ~~~ 1+1+1-64-64+125=0, 
\end{equation}
which is Solution (C) of Ref.~\cite{m21}.  It is based on the 
observation~\cite{mp09,ms15,cryz19,bccps20} that $(-1,-1,-1)$ may be 
replaced by $(5,-4,-4)$ as $B-L$ charges for gauge $B-L$ symmetry.  

The Higgs potential consists of six doublets $\Phi_{1,2},\eta_{1,2,3,4}$ 
(which are necessary to enforce the forms of charged-lepton and Dirac 
neutrino mass matrices to be discussed) and three singlets $\chi_{1,2,3}$ 
(which are necessary for masses of the dark fermions and the link between 
the $\Phi$ and $\eta$ doublets).  Their quadratic terms are such that 
$\Phi_{1,2},\chi_{1,2}$ have negative $\mu_{1,2,3,4}^2$, but 
$\eta_{1,2,3,4},\chi_3$ have large positive $m^2_{1,2,3,4,5}$.  The terms 
connecting them are
\begin{eqnarray}
&& f_1 \chi_1^2 \eta_1^\dagger \Phi_1 + f_2 \chi_1^2 \eta_2^\dagger \Phi_2 
+ f'_1 \chi_3 \chi_1^* \eta_1^\dagger \Phi_1 + f'_2 \chi_3 \chi_1^* 
\eta_2^\dagger \Phi_2 + \mu'_1 \chi_2 \eta_1^\dagger \Phi_1 + \mu'_2 
\chi_2 \eta_2^\dagger \Phi_2 
\nonumber \\ && + \mu_{12} \chi_2 \eta_1^\dagger \Phi_2 
+ \mu_{21} \chi_2 \eta_2^\dagger 
\Phi_1 + \mu_{31} \chi_2 (\eta_3^\dagger +\eta_4^\dagger) \Phi_1 + 
\mu_{32} \chi_2 (\eta_3^\dagger +\eta_4^\dagger) \Phi_2 \nonumber \\ 
&& + f_3 \chi_3^* \chi_1^3 + f_4 \chi_3^* \chi_2^2 \chi_1^* + 
\mu'_3 \chi_3^* \chi_2 \chi_1 + \mu'_4 \chi_2^* \chi_1^2 + H.c.,
\end{eqnarray}
where the $\mu_{12},\mu_{21},\mu_{31},\mu_{32}$ terms break $S_3$ softly. 
Let $\langle \phi^0_{1,2} \rangle = v_{1,2}$, 
$\langle \eta_{1,2,3,4}^0 \rangle = v'_{1,2,3,4}$, 
$\langle \chi_{1,2,3} \rangle = u_{1,2,3}$, then $v_{1,2},u_{1,2}$ obtain 
nonzero vacuum expectation values in the usual way at the breaking scales 
of $SU(2)_L \times U(1)_Y$ and $U(1)_D$ respectively, whereas 
$v'_{1,2,3,4},u_3$ are small because of the large positive $m^2_{1,2,3,4,5}$. 
Assuming that the soft breaking of $S_3$ preserves the interchange symmetry 
$\eta_3 \leftrightarrow \eta_4$, so that $m_3=m_4$ and $v'_3=v'_4$, the 
results are 
\begin{eqnarray}
&& v'_{1,2} \simeq {-f_{1,2} u_1^2 v_{1,2} - \mu'_{1,2} u_2 v_{1,2} 
- \mu_{12,21} u_2 v_{2,1} \over m^2_{1,2}}, \nonumber \\ 
&& v'_3 = v'_4 = \simeq {-\mu_{31} u_2 v_1 - \mu_{32} u_2 v_2 \over m_3^2}, 
~~~ u_3 \simeq {-f_3 u_1^3 - f_4 u_2^2 u_1 - \mu'_3 u_2 u_1 \over m_5^2}.
\end{eqnarray}

\noindent \underline{\it Dirac Neutrino Masses and Mixing}~:~
The $S_3$ representation used is that first proposed in Ref.~\cite{m91} 
and fully explained in Ref.~\cite{m04}.  Let $(a_1,a_2)$ and $(b_1,b_2)$ 
be doublets under $S_3$, then
\begin{equation}
a_1 b_2 + a_2 b_1 \sim 1, ~~~ a_1 b_2 - a_2 b_1 \sim 1', ~~~ 
(a_2 b_2, a_1 b_1) \sim 2.
\end{equation}
The structure of the charged-lepton mass matrix is determined thus by 
the Yukawa terms $y_1 e e^c \bar{\phi}_1^0$, 
$y_2 (\mu \mu^c + \tau \tau^c) \bar{\phi}_1^0$ and 
$y_3 (-\mu \mu^c + \tau \tau^c) \bar{\phi}_2^0$, so that the $3 \times 3$ 
mass matrix linking $(e,\mu,\tau)$ to $(e^c,\mu^c,\tau^c)$ is diagonal 
with $m_e = y_1 v_1$, $m_\mu = y_2 v_1 - y_3 v_2$, 
$m_\tau = y_2 v_1 + y_3 v_2$.

There are only two singlet neutrinos $\nu^c_{2,3}$ which couple to 
$(\nu_e,\nu_\mu,\nu_\tau)$ through $\eta^0_{1,2,3,4}$.  One linear 
combination of the three neutrinos must then be massless.  For calculational 
convenience, $\nu^c_1$ may be added, so that the $3 \times 3$ Dirac mass 
matrix is of the form
\begin{equation}
{\cal M}_\nu = \pmatrix{0 & a & b \cr 0 & c & -d \cr 0 & c & d},
\end{equation}
where the (12) entry comes from $v'_1$, the (13) entry comes from $v'_2$, 
the (22) and (32) entries are the same because they come from 
$(\mu \eta^0_4 + \tau \eta_3^0)\nu^c_2$, whereas the (23) and (33) entries 
come from $(-\mu \eta^0_4 + \tau \eta_3^0)\nu_3^c$.
 
The neutrino mixing matrix is then obtained by diagonalizing 
\begin{equation}
{\cal M}_\nu {\cal M}_\nu^\dagger = \pmatrix{ |a|^2 + |b|^2 & ac^*-bd^* & 
ac^* + bd^* \cr a^*c-b^*d & |c|^2+|d|^2 & |c|^2-|d|^2 \cr a^*c + b^*d & 
|c|^2 - |d|^2 & |c|^2+|d|^2}.
\end{equation}
Assuming that $|a|^2|b|^2 << (2|d|^2+|b|^2)(2|c|^2+|a|^2)$, the eigenvalues 
are 
\begin{equation}
m_{\nu_1} = 0, ~~~ m_{\nu_2}^2 = 2|c|^2 + |a|^2, ~~~ 
m_{\nu_3}^2 = 2|d|^2 + |b|^2.
\end{equation}
Let $b/d=i \sqrt{2} s_{13}/c_{13}$, then 
\begin{equation}
\nu_3 = i s_{13} \nu_e - {1 \over \sqrt{2}} c_{13} \nu_\mu + 
{1 \over \sqrt{2}} c_{13} \nu_\tau.
\end{equation}
Let $a/c=\sqrt{2} c_{13} s_{12}/c_{12}$, then 
\begin{equation}
\nu_2 = s_{12} c_{13} \nu_e + {1 \over \sqrt{2}} (c_{12} - 
i s_{12} s_{13}) \nu_\mu + {1 \over \sqrt{2}}( c_{12} + i s_{12} s_{13}) 
\nu_\tau.
\end{equation}
With these choices, the massless eigenstate is automatically
\begin{equation}
\nu_1 = c_{12} c_{13} \nu_e + {1 \over \sqrt{2}} (-s_{12} - 
i c_{12} s_{13}) \nu_\mu + {1 \over \sqrt{2}}(- s_{12} + i c_{12} s_{13}) 
\nu_\tau.
\end{equation}
In other words, a completely realistic neutrino mixing scenario dubbed 
cobimaximal~\cite{m16-1} with $\theta_{23} = \pi/4$ and Dirac CP phase 
$\delta = -\pi/2$ is possible with Eq.~(5).  Numerically, using the most 
recent world averages~\cite{pdg20}
\begin{equation}
m^2_{32} = 2.453 \times 10^{-3}~{\rm eV}^2, ~~~ m^2_{21} = 7.53 \times 
10^{-5}~{\rm eV}^2, ~~ s_{13}^2 = 0.0218, ~~~ s_{12}^2 = 0.307,
\end{equation}
the values
\begin{equation}
d = 0.035~{\rm eV}, ~~~ c = 0.0051~{\rm eV}, ~~~ b/d = 0.21 i, ~~~ 
a/c = 0.93
\end{equation}
are obtained.

\noindent \underline{\it Deviation from Cobimaximal Mixing}~:~
Using $v'_3=v'_4$, the correlation of $\delta = -\pi/2$ to $\theta_{23}=\pi/4$ 
has been obtained with Eq.~(5).  Since the most recent data~\cite{pdg20} 
favors $\theta_{23} > \pi/4$, a modification of Eq.~(5) is studied to 
see how $\delta$ changes numerically with $\theta_{23}$.  Let 
\begin{equation}
{\cal M}_\nu = \pmatrix{0 & a & b \cr 0 & (1+\epsilon)c & -(1+\epsilon)d 
\cr 0 & (1-\epsilon)c & (1-\epsilon)d},
\end{equation}
then
\begin{equation}
{\cal M}_\nu {\cal M}_\nu^\dagger = \pmatrix{ |a|^2 + |b|^2 & (1+\epsilon)
(ac^*-bd^*) & (1-\epsilon)(ac^* + bd^*) \cr (1+\epsilon)(a^*c-b^*d) 
& (1+\epsilon)^2(|c|^2+|d|^2) & (1-\epsilon^2)(|c|^2-|d|^2) \cr (1-\epsilon)
(a^*c + b^*d) & (1-\epsilon^2)(|c|^2 - |d|^2) & (1-\epsilon)^2(|c|^2+|d|^2)}.
\end{equation}
To first order in $\epsilon$, the mass eigenvalues are the same, with 
the following changes in the mixing parameters: 
\begin{eqnarray}
&& s_{23} = -{1 \over \sqrt{2}} (1+\epsilon), ~~~ c_{23} = {1 \over \sqrt{2}} 
(1-\epsilon), ~~~ e^{-i\delta} = i e^{i\theta'}, \\ 
&& {b \over d} = i \sqrt{2} {s_{13} \over c_{13}} e^{i \theta'}, ~~~ 
{c \over a} = {c_{12} - 2i\epsilon s_{12} s_{13} \over \sqrt{2} 
s_{12} c_{13}}, ~~~ \theta' = {2 \epsilon |c|^2 \over |d|^2} 
{s_{12} c_{13}^2 \over c_{12} s_{13}}.
\end{eqnarray}
Using the world average~\cite{pdg20} $s^2_{23} = 0.545$, the deviation 
from cobimaximal mixing is then
\begin{equation}
\epsilon = 0.044, ~~~ \theta' = 8.24 \times 10^{-3}.
\end{equation}

\noindent \underline{\it Dark Sector}~:~
The dark sector consists of four fermion singlets $\psi_{1,2,3} \sim 1$ 
and $\zeta \sim 5$.  Because of the chosen scalars $\chi_1 \sim 2$, 
$\chi_2 \sim 4$ and $\chi_3 \sim 6$, there is no connection to the two 
singlet neutrinos $\nu^c_{2,3} \sim -4$.  Hence $\psi_{1,2,3},\zeta$ 
may be considered odd under an induced $Z_2$ symmetry which stabilizes 
the lightest among them as dark matter.  The $4 \times 4$ Majorana mass 
matrix spanning $(\zeta,\psi_{1,2,3})$ is of the form
\begin{equation}
{\cal M}_{\zeta,\psi} = \pmatrix{ 0 & h'_1 u_3 & h'_2 u_3 & h'_3 u_3 \cr 
h'_1 u_3 & h_1 u_1 & 0 & 0 \cr h'_2 u_3 & 0 & h_2 u_1 & 0 \cr h'_3 u_3 & 
0 & 0 & h_3 u_1}.
\end{equation}
Recalling that $u_3 << u_1$ from Eq.~(3), it is clear that $\zeta$ gets 
a very small mass, i.e.
\begin{equation}
m_\zeta = - {{h'_1}^2 u_3^2 \over h_1 u_1} - {{h'_2}^2 u_3^2 \over h_2 u_1} 
- {{h'_3}^2 u_3^2 \over h_3 u_1}.
\end{equation}

Since $\chi_3$ has large and positive $m_5^2$ so that $u_3$ is very small, 
the breaking of $U(1)_D$ is mainly through $\chi_{1,2}$.  The relevant 
part of the Higgs potential is then
\begin{eqnarray}
V &=& -\mu_3^2 \chi_1^* \chi_1 - \mu_4^2 \chi_2^* \chi_2 + [\mu'_4 \chi_2^* 
\chi_1^2 + H.c.] \nonumber \\ 
&+& {1 \over 2} \lambda_1 (\chi_1^* \chi_1)^2 + {1 \over 2} \lambda_2 
(\chi_2^* \chi_2)^2 + \lambda_3 (\chi_1^* \chi_1)(\chi_2^* \chi_2).
\end{eqnarray}
Let $H_{1,2} = \sqrt{2} Re(\chi_{1,2})$, then the $2 \times 2$ mass-squared 
matrix spanning $H_{1,2}$ is
\begin{equation}
{\cal M}^2_H = \pmatrix{ 2 \lambda_1 u_1^2 & 2\lambda_3 u_1 u_2 + 2 \mu'_4 u_1 
\cr 2\lambda_3 u_1 u_2 + 2 \mu'_4 u_1 & 2\lambda_2 u_2^2 - \mu'_4 u_1^2/u_2}.
\end{equation}
Let $H_2$ be the lighter, with mixing $\theta$ to $H_1$.  Now $H_2$ does not 
couple to $\psi \psi$, but $H_1$ does and through $\psi-\zeta$ mixing to 
$\zeta \zeta$ with Yukawa coupling $y_{H} = m_\zeta/2 \sqrt{2} u_1$.  
From Eqs.~(3) and (19), it is clear that $m_\zeta << u_1$, hence 
$y_{H}$ is very much suppressed.

Consider now the very light Majorana fermion $\zeta$ as dark matter.  It has 
gauge interactions, but if the reheat temperature of the Universe is much 
below the mass of the $U(1)_D$ gauge boson as well as $m_{H_1}$, then it 
interacts only very feebly through $H_2$.  Its production mechanism in 
the early Universe is freeze-in~\cite{hjmw10} by $H_2$ decay before the 
latter decouples from the thermal bath.  The decay rate of 
$H_2 \to \zeta \zeta$ is
\begin{equation}
\Gamma_{H_2} = {y_{H}^2 \theta^2 m_{H_2} \over 8 \pi} \sqrt{1-4r^2} (1-2r^2),
\end{equation}
where $r= m_\zeta/m_{H_2}$.  For $r << 1$, the correct relic abundance 
is obtained for~\cite{ac13}
\begin{equation}
y_{H} \theta \sim 10^{-12} r^{-1/2}.
\end{equation}
This translates to
\begin{equation}
{m_\zeta \theta \over (u_1^2m_{H_2})^{1/3}} \sim 2 \times 10^{-8},
\end{equation}
which may be satisfied for example with $u_1 = 10^7$ GeV, $m_{H_2} = 600$ GeV, 
$\theta = 0.1$, and $m_\zeta = 80$ MeV.  The decoupling temperature for 
$\zeta$ is roughly $T \sim 1~{\rm MeV} (m_{Z_D}/m_Z)^{4/3} = 5.2$ TeV.  
This analysis follows that of Ref.~\cite{m19}, where the SM Higgs decay to 
a light seesaw dark fermion from the decomposition of 
$SU(10) \to SU(5) \times U(1)_\chi$.

\noindent \underline{\it Concluding Remarks}~:~
The right-handed neutrino $\nu_R$ has been proposed as the link~\cite{m21} 
to dark matter by having it transform under a new $U(1)_D$ gauge symmetry. 
The small Dirac neutrino masses are enforced by a seesaw mechanism proposed 
in Ref.~\cite{m01} using new Higgs doublets also transforming under $U(1)_D$. 
With the help of the non-Abelian discrete symmetry $S_3$, it is shown how a 
realistic neutrino mixing matrix may be obtained which is approximately 
cobimaximal~\cite{m16-1}.

After the spontaneous breaking of $U(1)_D$, a dark parity remains for four 
Majorana fermions, the lightest of which has a seesaw mass.  It is suitable 
as freeze-in dark matter with its relic abundance coming from the decay of 
the $U(1)_D$ Higgs boson.

\noindent \underline{\it Acknowledgement}~:~
This work was supported in part by the U.~S.~Department of Energy Grant 
No. DE-SC0008541.

\bibliographystyle{unsrt}

\begin{thebibliography}{99}
\bibitem{m21} E. Ma, arXiv:2101.12138 [hep-ph].
\bibitem{m01} E. Ma, Phys. Rev. Lett. {\bf 86}, 2502 (2001).
\bibitem{m98} E. Ma, Phys. Rev. Lett. {\bf 81}, 1171 (1998).
\bibitem{ms98} E. Ma and U. Sarkar, Phys. Rev. Lett. {\bf 80}, 5716 (1998).
\bibitem{glr09} W. Grimus, L. Lavoura, and B. Radovcic, Phys. Lett. 
{\bf B674}, 117 (2009).
\bibitem{m16} E. Ma, Phys. Rev. {\bf D94}, 031701(R) (2016).
\bibitem{mp09} J. C. Montero and V. Pleitez, Phys. Lett. {\bf B675}, 64 (2009).
\bibitem{ms15} E. Ma and R. Srivastava, Phys. Lett. {\bf B741}, 217 (2015).
\bibitem{cryz19} J. Calle, D. Restrepo, C. E. Yaguna, and O. Zapata, Phys. 
Rev. {\bf D99}, 075008 (2019).
\bibitem{bccps20} C. Bonilla, S. Centelles Chulia, R. Cepedello, E. Peinado, 
and R. Srivastava, Phys. Rev. {\bf D101}, 033011 (2020).
\bibitem{m91} E. Ma, Phys. Rev. {\bf D43}, 2761(R) (1991). 
\bibitem{m04} E. Ma, Fuji Lectures (SI 2004), hep-ph/0409075 (2004).
\bibitem{m16-1} E. Ma, Phys. Lett. {\bf B752}, 198 (2016).
\bibitem{pdg20} P. A. Zyla {\it et al.} (Particle Data Group), Prog. 
Theor. Exp. Phys. {\bf 2020}, 083C01 (2020). 
\bibitem{hjmw10} L. J. Hall, K. Jedamzik, J. March-Russell, and S. M. West, 
JHEP {\bf 1003}, 080 (2010).
\bibitem{ac13} G. Arcadi and L. Covi, JCAP {\bf 1308}, 005 (2013).
\bibitem{m19} E. Ma, LHEP {\bf 2.1}, 103 (2019) [arXiv:1810.06506].
\end{thebibliography}

\end{document}